\documentclass[pdflatex,sn-mathphys-num]{sn-jnl}

\usepackage[utf8]{inputenc}
\usepackage[T1]{fontenc}
\usepackage[english]{babel}

\usepackage{amsmath,amssymb,amsfonts,bm,dsfont}
\usepackage{braket}
\usepackage{empheq}

\usepackage{graphicx}
\usepackage{caption}
\usepackage{subcaption}

\usepackage{booktabs}
\usepackage{multirow}
\usepackage{microtype}
\usepackage{ulem} 

\begin{document}
\title[Quantum Optimal Control in an Antiferromagnetic XXZ Chain]{Quantum Optimal Control of Many-Body States in an Antiferromagnetic XXZ Chain with a Single Local Magnetic Field}

\author[1]{\fnm{Nicol\'as} \sur{Rojo}}

\author[1]{\fnm{Ignacio R.} \sur{Sol\'a}}

\author*[1]{\fnm{Juan J.} \sur{Omiste}}\email{jjomiste@ucm.es}

\affil*[1]{\orgdiv{Departamento de Qu\'imica F\'isica}, \orgname{Universidad Complutense de Madrid}, \orgaddress{\postcode{28040}, \city{Madrid}, \country{Spain}}}

\abstract{We investigate the quantum optimal control of an antiferromagnetic XXZ spin chain with open boundary conditions, driven solely by a local time-dependent magnetic field applied to a single terminal spin. By optimizing the control field via Krotov's algorithm, we evaluate the feasibility of preparing multi-particle many-body states, including highly polarized Greenberger-Horne-Zeilinger (GHZ)-type states and study the mechanisms by which the control is applied. We demonstrate that global spin-flip symmetry restricts reachable state spaces, strictly confining dynamics within the parity sector of the initial ground state and requiring parity-adapted linear combinations to avoid fundamental population upper bounds. Beyond symmetry constraints, driven population transfer does not occur via direct multi-spin flips, but proceeds stepwise across intermediate magnetization sectors ($|S_{\mathrm{tot}}^z| = 0 \to 1 \to 2 \to 3$) mediated by resonant frequency components matching the internal energy gaps. To overcome optimization stagnation caused by the small overlap mechanism in gradient updates, we introduce a four-stage sequential target protocol and compared it to spectrally filtered, time-localized frequency seed pulses. Combining frequency-tailored seed fields with an extended propagation time window yields GHZ state preparation fidelities exceeding $0.927$ establishing that a single local actuator is sufficient for high-fidelity multipartite state engineering in interacting spin networks.}
\keywords{Quantum optimal control, XXZ spin chain, Krotov algorithm, GHZ states, Many-body dynamics}

\maketitle

\section{Introduction}
\label{sec:introduction}

In quantum physics and chemistry, Quantum Optimal Control Theory (QOCT) offers a powerful tool to manipulate quantum states, observable expectation values, and unitary transformations~\cite{werschnik2007}. Initially developed to steer wave packet dynamics and select molecular reaction pathways~\cite{tannor1985,peirce1988,tannor1992,zhu1998}, QOCT now is applied to quantum information processing~\cite{palao2002}, quantum sensing~\cite{poggiali2018}, and circuit QED engineering~\cite{motzoi2009,goerz2017}. Control mechanisms like STIRAP~\cite{bergmann1998,vitanov2017} and pulse shaping~\cite{shapiro2003,sola2018} offer physical intuition, while algorithms such as Krotov optimization provide systematic field updating~\cite{krotov1993,reich2012}.

In quantum computing, QOCT serves as an essential framework for compiling high-fidelity quantum gates devices~\cite{sauvage2022}. Algorithms such as GRAPE~\cite{khaneja2005} optimize pulse sequences while incorporating physical constraints, including bandwidth limitations, leakage to non-computational states, and open-system dissipation. Beyond single-qubit manipulations, quantum entanglement serves as the foundational resource for quantum information processing, enabling non-local correlations without classical analogs~\cite{horodecki2009}. In multipartite systems, Greenberger--Horne--Zeilinger (GHZ) states is a $N$-qubit maximally entangled states~\cite{greenberger1989}. While carrying perfect global correlations, GHZ states exhibit an extreme form of fragility: measuring or tracing out a single particle completely collapses the quantum entanglement, reducing the remaining sub-system to a fully separable classical mixture. Despite this susceptibility to local particle loss, their collective non-local structure makes GHZ states essential across quantum protocols. In particular, their global correlations supports multi-party communication frameworks such as Quantum Secret Sharing (QSS)~\cite{hillery1999}, while their high phase sensitivity enables quantum metrology applications to surpass the Standard Quantum Limit and achieve Heisenberg-scaling precision~\cite{giovannetti2004, degen2017}.

Applying QOCT to many-body systems, like spin chains, is central to quantum communications~\cite{bose2003,bose2008} and multipartite entanglement generation~\cite{greenberger1990,hillery1999,giovannetti2004,giovannetti2006}. A critical challenge arises when independent control of every site is unfeasible, requiring a single or few local fields to drive the global spin array through internal exchange interactions~\cite{heule2010,heule2011,castro2025,schirmer2001,schirmer2003}, making mandatory to determine the degree of controllability~\cite{RamakrishnaPRA95,TuriniciJPA03} in many-body quantum networks and to identify the minimal physical resources required to achieve high-fidelity targets~\cite{Burgarth2009,Wang2016}. In addition, for an $N$-body system, applying $N$ independent control fields is often technologically prohibitive and introduces additional channels for decoherence. Exploiting the internal dynamics mediated by spin-spin exchange couplings allows a single or few local boundary controls to indirectly steer the entire spin array~\cite{Burgarth2007}. While Hamiltonian symmetries (e.g., total magnetization conservation) restrict global Hilbert space controllability into distinct invariant subspaces, localized fields remain fully capable of achieving near-unit state preparation fidelity within a targeted subspace~\cite{Wang2016}.

In this investigation we tackle the control of the global 1D spin chain by acting locally on a single site. Specifically, we focus on the state preparation in a $N=6$ spin antiferromagnetic XXZ chain with open boundaries controlled by a transverse magnetic field applied to the first site. In many-body systems with scarce control resources, as this work, it is essential to design strategies to find the optimal fields whenever the distance from the initial state to the target state is so large (in the energy representation) and involves so many quantum jumps, that the transition is highly unlikely and the optimal algorithm is deemed to fail, particularly at short propagation times. 
We explore the symmetry constraints of the system and propose two strategies to guide the optimization: by including sub-target states at intermediate times, and by seeding the optimal field with convenient parameterizations, using spectral analysis to build high-fidelity initialization pulses for GHZ-type target states.

The structure of this work is as follows: in Sec~\ref{sec:theory_methods} we describe the Hamiltonian of the system under study and explain its symmetrical properties. The computational methods used are briefly described. Then, in Sec.~\ref{sec:results} we introduce the multistage protocol applied to achieve the target state. Using this result, we derive an efficient transition process based on an appropriate sequence of fixed frequency pulses. Finally, in Sec.~\ref{sec:conclusions} we summarize the main findings of this work and present the application of the multistage protocol to other systems of interest.

\section{Theoretical Model and Methodology}
\label{sec:theory_methods}

\subsection{Model and Symmetry Considerations}
\label{sec:model_symmetry}
We consider a 1D spin-1/2 XXZ chain of length $N=6$ with nearest-neighbor interactions in the presence of a control field. The total time-dependent Hamiltonian reads as $\hat{H}(t) = \hat{H}_0 + \varepsilon(t)\hat{H}_c$. The field-free Hamiltonian is~\cite{heule2010,heule2011,Mattis_2006}
\begin{equation}
\hat{H}_0 = \sum_{k=1}^{5} \left[ \frac{J}{2}\left(S_k^+ S_{k+1}^- + S_k^- S_{k+1}^+\right) + \Delta S_k^z S_{k+1}^z \right],
\label{eq:H0}
\end{equation}
where $J$ is the exchange coupling, $\Delta$ is the anisotropy parameter, and $S_k^\pm = S_k^x \pm i S_k^y$. 
Note that $H_0$ conserves the total magnetization along the $z$-axis, $[ \hat{H}_0, \sum_i S_k^z ] = 0$, therefore, the number of particles pointing to opposite directions is preserved~\cite{castro2025}, defining different sectors of states of the Hamiltonian. Furthermore, it possesses global spin-flip symmetry under $\hat{P}_x = \prod_{i=1}^N (2 S_i^x)$. 
\\
The system is driven by a local transverse magnetic field $\varepsilon(t)$ acting on the first site ($k=1$), described by~\cite{heule2010,Mattis_2006}
\begin{equation}
\hat{H}_c =- S_1^x = -\frac{1}{2}\left(S_1^+ + S_1^-\right).
\label{eq:Hc}
\end{equation}
\\
Because the local control operator $\hat{H}_c = -S_1^x$ commutes with $\hat{P}_x$, the total Hamiltonian preserves parity
\begin{equation}
[\hat{H}(t), \hat{P}_x] = 0.
\end{equation}
Note that the coupling with $H_c(t)$ is non-zero if the states involved fulfill $\Delta |S^z_{tot}|=1$. 

Regarding the computational performance, the numerical basis set is built up from the symmetrization of $\ket{c}$, defined as the direct product of single particle states with well-defined projection of the magnetization along the $Z$-axis, i.e., $S^z$. We label particle states with $\ket{1}$ for spin-up and $\ket{0}$ for spin down. 

\subsection{Optimal Quantum Control}
To prepare a target state $\ket{\psi_{\mathrm{tar}}}$, we define the objective projector $\hat{O} = \ket{\psi_{\mathrm{tar}}}\bra{\psi_{\mathrm{tar}}}$. The functional to maximize is~\cite{tannor1992,werschnik2007}
\begin{equation}
J = \left|\braket{\psi_{\mathrm{tar}}|\psi(T)}\right|^2 - \int_0^T \frac{\alpha}{S(t)} \varepsilon^2(t)\, dt,
\label{eq:functional}
\end{equation}
where $J_T = \left|\braket{\psi_{\mathrm{tar}}|\psi(T)}\right|^2$ is the terminal target population, $\alpha > 0$ is the penalty parameter, and $S(t)$ is a pulse envelope shape function.

Using Krotov's method~\cite{krotov1993,reich2012}, field updates at iteration $i+1$ are constructed self-consistently:
\begin{equation}
\varepsilon^{(i+1)}(t) = \frac{S(t)}{\alpha} \mathrm{Im} \left[ \bra{\chi^{(i)}(t)} \hat{H}_c \ket{\psi^{(i+1)}(t)} \right],
\label{eq:krotov_update}
\end{equation}
where the forward state evolves from $\ket{\psi^{(i+1)}(0)} = \ket{E_1}$, and the backward costate $\chi^{(i)}(t)$, which is defined as a dynamical Lagrange multiplier that constrains the dynamics to follow the time-dependent Schr\"odinger equation (TDSE), evolves from the terminal condition $\ket{\chi^{(i)}(T)} = \hat{O}\ket{\psi^{(i)}(T)}$. The set of equations are completed with the dynamic evolution of $\ket{\psi^{(i+1)}(t)} = U(t,0;\epsilon^{(i+1)}(t)) \ket{\psi(0)}$ and $\ket{\chi^{(i)}(t)} = U^{-1}(T,t;\epsilon^{(i)}(t)) \ket{\chi^{(i)}(T)}$, where $U(t,0;\epsilon^{(i+1)}(t))$ is the time evolution operator, which depends on the optimal field. Fast feedback control in the algorithm is achieved by propagating $\ket{\chi^{(i)}(t)}$ backward in time, with pulse $\epsilon^{(i)}(t)$ and $\ket{\psi^{(i+1)}(t)}$ forward in time, with pulse $\epsilon^{(i+1)}(t)$. The numerical simulations in this work are performed with a total propagation time of $T = 100$ in dimensionless units, discretized into $N_t = 1.5 \times 10^6$ time steps. The duration of each time step is therefore $\Delta t \approx 6.67 \times 10^{-5}$. The numerical integration of the associated time-dependent Schr\"odinger equations is performed using a fourth-order Runge-Kutta scheme (RK4) as implemented within the~\texttt{QOCY} package~\cite{qoc_routines_github}.

\section{Results}
\label{sec:results}
In this section, we aim to maximize the population of a given state of the chain at a given measuring time.  

First, we briefly describe the system and then propose several strategies within Quantum Optimal Control theory to achieve the desired population transfer.

\subsection{Eigenstates and computational considerations}
\label{sec:eigenstates_compuational}
Let us first describe the eigenstates and the implications of the symmetry in both the structure of the energy levels and the dynamics. In this work, we describe an anti-ferromagnet with a non degenerate ground state by fixing $\Delta = 2$ and $J = 1$.

The field-free Hamiltonian $\hat{H}_0$ yields a non-degenerate ground state $\ket{E_1}$ in the $S_{\mathrm{tot}}^z = 0$ sector. Because of parity conservation under $\hat{P}_x$, targets must be constructed as parity-adapted superpositions. For any computational state $\ket{c}$, the appropriate target is $\frac{1}{\sqrt{2}}(\ket{c} - \ket{\bar{c}})$, where $\ket{\bar{c}}$ represents global spin inversion ($\ket{0} \leftrightarrow \ket{1}$). Therefore, the dynamics are strictly confined within the parity subspace populated by the initial ground state $\ket{E_1}$. Target states outside this parity sector are dynamically inaccessible. Maximizing the population of non-parity-adapted single basis configurations (such as $\ket{101010}$) results in an upper bound $J_T \le 0.5$. 
 This is a clear signature that the system is uncontrollable~\cite{RamakrishnaPRA95,schirmer2001,TuriniciJPA03}, but connected~\cite{WuPRA04}. 
\begin{figure}[htbp]
\centering
\includegraphics[width=\textwidth]{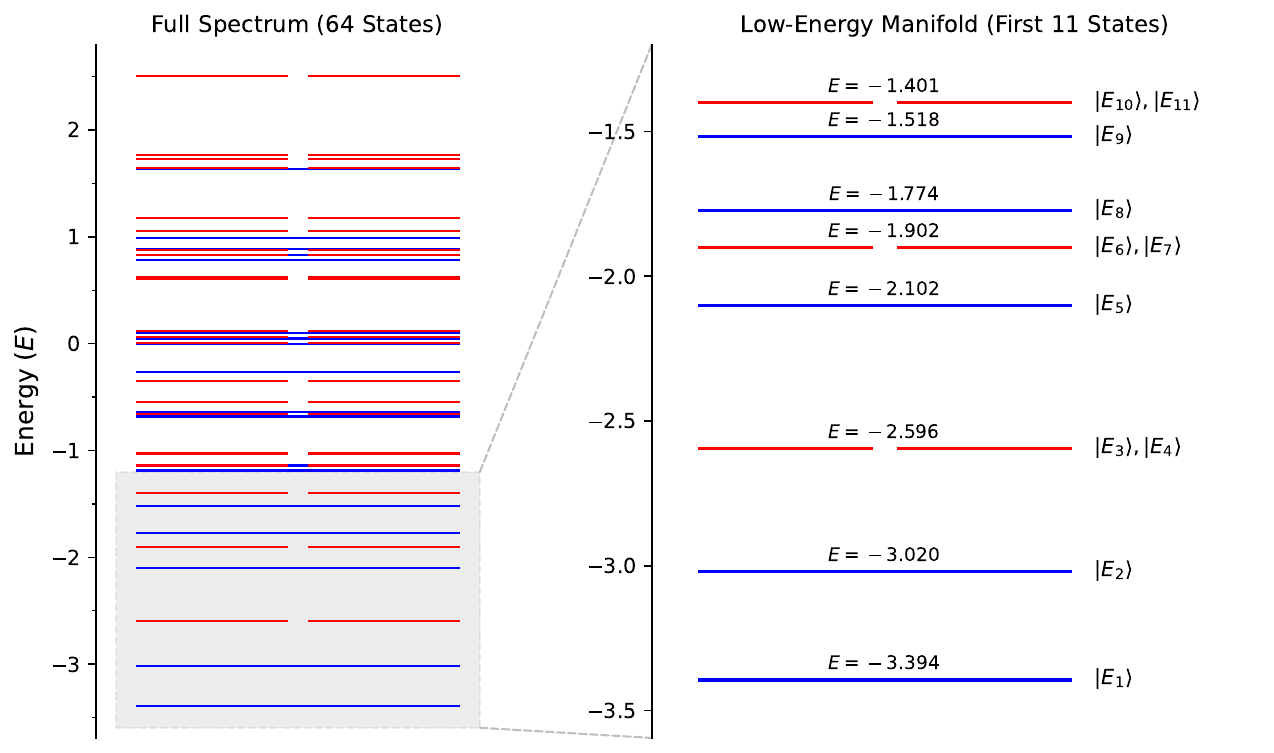}
\caption{Energy spectrum of the field-free Hamiltonian for the $N=6$ spin chain with $\Delta=2$ and $J=1$. The inset shows the low lying energy levels.}
\label{fig:combined_spectrum}
\end{figure}
In Fig.~\ref{fig:combined_spectrum} the full energy spectrum of the field-free Hamiltonian $\hat{H}_0$ is displayed. The non degenerate energy levels are represented in blue, whereas red lines indicate pairs of degenerate levels of opposite symmetry under spin-flip. The ground state $|E_1\rangle$ is non degenerate, with energy $E_1=-3.394$. The expansion in the basis set is
\begin{equation}
\begin{split}
    |E_1\rangle =\; & \phantom{+}0.5326 \left( |010101\rangle - |101010\rangle \right) \\
                & - 0.2758 \left( |010110\rangle - |101001\rangle \right) \\
                & + 0.2758 \left( |011010\rangle - |100101\rangle \right) \\
                & - 0.1873 \left( |011001\rangle - |100110\rangle \right) \\
                & - 0.1070 \left( |001101\rangle - |110010\rangle \right) \\
                & - 0.1070 \left( |010011\rangle - |101100\rangle \right) \\
                & + 0.0491 \left( |001110\rangle - |110001\rangle \right) \\
                & - 0.0491 \left( |011100\rangle - |100011\rangle \right) \\
                & + 0.0376 \left( |001011\rangle - |110100\rangle \right) \\
                & - 0.0038 \left( |000111\rangle - |111000\rangle \right)
\end{split}
\label{eq:ground_state_full}
\end{equation}
As expected from the symmetries of \(\hat{H}_0\) discussed in Sec.~\ref{sec:model_symmetry}, total magnetization of the ground state is well-defined with $S_{\mathrm{tot}}^z=0$. The reflection symmetry with respect to the center of the chain, and the global spin inversion symmetry, imply that the exchange of any spin orientation is accompanied by a sign change in the wave function. This relative sign between globally inverted configurations obeys Marshall's sign rule~\cite{Voigt1997} and is given by the factor $(-1)^{N/2}$.
\subsection{Basis State Optimizations and Small Overlap Mechanism}
\begin{figure}
    \centering
    \includegraphics[width=0.85\linewidth]{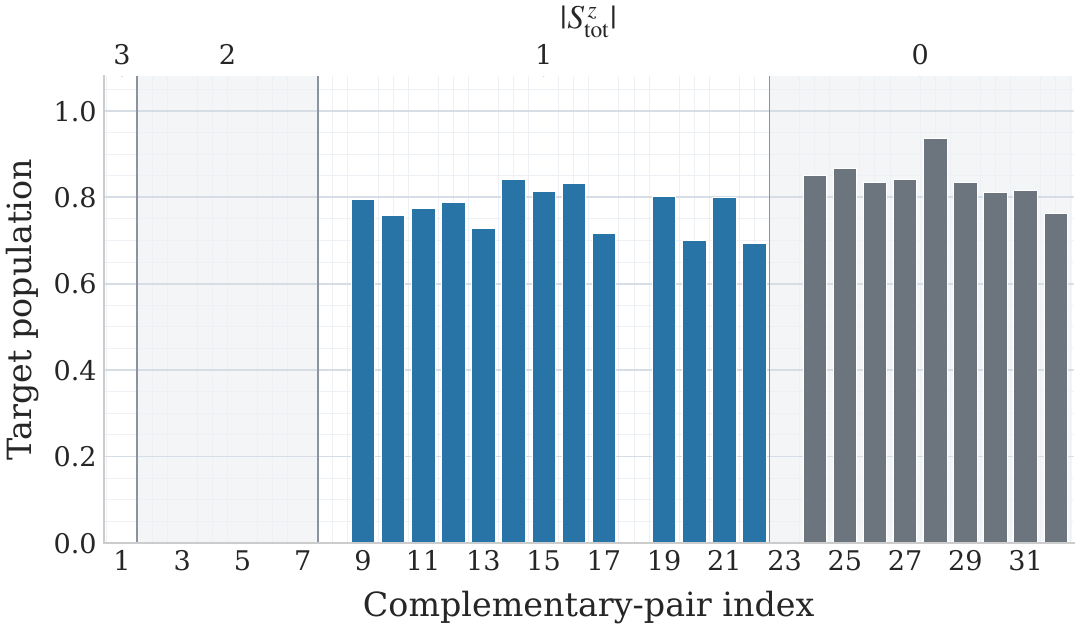}
    \caption{Target state population obtained taking the function in Eq.~\eqref{eq:hahn_seed} as seed for the control field. Note that the complementary-pair index correspond to the computational labeling of the symmetrized basis set containing the ground state, for which we only show the population of half the states, the other half being strictly zero by symmetry}.
    \label{fig:target_population_hahn}
\end{figure}

Taking the absolute ground state as the initial wave function, we explore the maximization of the population for every element of the basis set, formed by the 32 symmetrized functions characterized by the same symmetrical properties as the ground state.
\\
We first set the guess field,~$\varepsilon^{(0)}(t)$, used as a seed zeroth iteration of the Krotov algorithm
\begin{equation}
\label{eq:hahn_seed}
\varepsilon^{(0)}(t) = 
    \begin{cases}
        \cos^2\left(\frac{\pi (t-5)}{10}\right) &\text{for } t \in [0, 10],\\
        \\
        0,&\text{otherwise}.
    \end{cases}
\end{equation}
\\
In Fig.~\ref{fig:target_population_hahn}, we display the target population obtained for each element of the basis set. We observe that high target populations ($J_T > 0.7 - 0.9$) are systematically achieved for eigenstate characterized by $|S_{\mathrm{tot}}^z| = 0$ and $1$. 
However, targets with $|S_{\mathrm{tot}}^z| = 2, 3$ yield $J_T \approx 0$. This is due to the weak optimal field being too weak. Following Eq.~\eqref{eq:krotov_update}, $\varepsilon(t)$ is given by the overlap of the costate, $\ket{\chi(t)}$, with the propagated wave packet in the first iteration. Because the overlap is small, the update rule prioritizes minimizing the field penalty over maximizing the final population, which leads to a further decrease in the field amplitude, causing the vanishing of the final population of the target eigenstate. As we highlighted in Sec.~\ref{sec:eigenstates_compuational}, the control Hamiltonian only connects states fulfilling $\Delta|S^{z}_\text{tot}|=1$. Hence, populating other contributions in the costate, $\ket{\chi(T)}=\hat{O}\ket{\psi(T)}$, would require a more sophisticated seed for the field, so that states with $|S_{\mathrm{tot}}^z| = 2, 3$ are relevant in the costate at the first iteration. This would guaranteed a moderate overlap in Eq.~\eqref{eq:krotov_update}. 

\subsection{Sequential Protocol for High-Magnetization Targets}
\label{sec:sequential_protocol}

To overcome the small-overlap bottleneck, we introduce a multi-stage seed-design algorithm that prevents the update overlap in Eq.~\eqref{eq:krotov_update} from vanishing. With this procedure, we aim to design a seed field that is able to increase the overlap of the costate and the wave packet during the propagation. The protocol relies on two core principles: first, each stage uses the converged control field from the preceding step as its initial seed; second, the intermediate target states are selected sequentially along allowed excitation pathways driven by the control field. The sketch of the algorithm is
\\
\begin{verbatim}
Multi-Stage Sequential Optimization
--------------------------------------------------------------------------------
Inputs:
  - Initial state: |E_1>
  - Ordered sequence of intermediate targets: {|psi_1>, |psi_2>, ..., |psi_M>}
  - Initial seed field: epsilon_0(t)

Initialization:
  Set current_seed = epsilon_0(t)

For stage m = 1 to M do:
  1. Define target projector: O_m = |psi_m><psi_m|
  2. Set field initialization: epsilon^(0)(t) = current_seed(t)
  3. Run Krotov Optimization toward |psi_m>.
  4. Extract optimized field: epsilon_opt_m(t)
  5. Update seed for next stage: current_seed(t) = epsilon_opt_m(t)
End For

Output:
  Final optimized control field driving the system to target state |psi_M>
--------------------------------------------------------------------------------
\end{verbatim}

As a benchmark of the method, we take as target, the GHZ state, $\ket{\psi_{\mathrm{GHZ}}} = \frac{1}{\sqrt{2}}(\ket{111111} - \ket{000000})$, which is an essential state for Quantum Information protocols and entanglement preparation. To circumvent small-overlap stagnation, we employ a sequential stage-wise protocol where the target state is shifted progressively across intermediate magnetization sectors:
\begin{enumerate}
    \item Stage 1: $\frac{1}{\sqrt{2}}(\ket{110101} - \ket{001010})$ ($|S_{\mathrm{tot}}^z|=1$).
    \item Stage 2: $\frac{1}{\sqrt{2}}(\ket{111001} - \ket{000110})$ ($|S_{\mathrm{tot}}^z|=1$).
    \item Stage 3: $\frac{1}{\sqrt{2}}(\ket{111110} - \ket{000001})$ ($|S_{\mathrm{tot}}^z|=2$).
    \item Stage 4: $\frac{1}{\sqrt{2}}(\ket{111111} - \ket{000000})$ ($|S_{\mathrm{tot}}^z|=3$).
\end{enumerate}

\begin{figure}[htbp]
\centering
    \begin{subfigure}[b]{0.48\textwidth}
        \includegraphics[width=\textwidth]{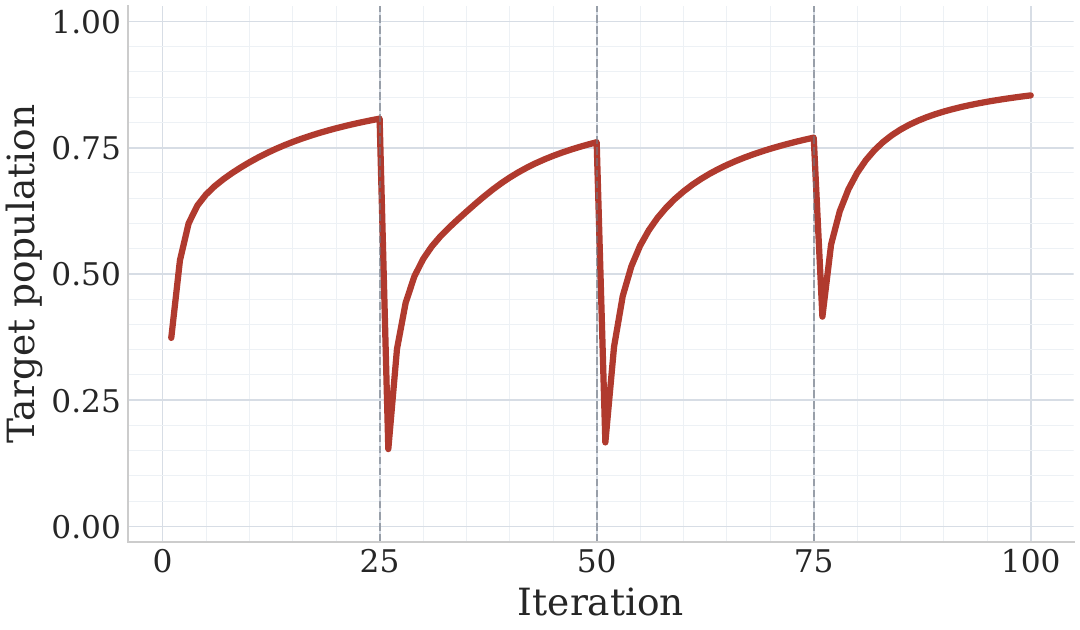}
        \caption{$\alpha = 10^{-2}$, 25 iterations per stage}
        \label{fig:sweep_success}
    \end{subfigure}
    \hfill
    \begin{subfigure}[b]{0.48\textwidth}
        \includegraphics[width=\textwidth]{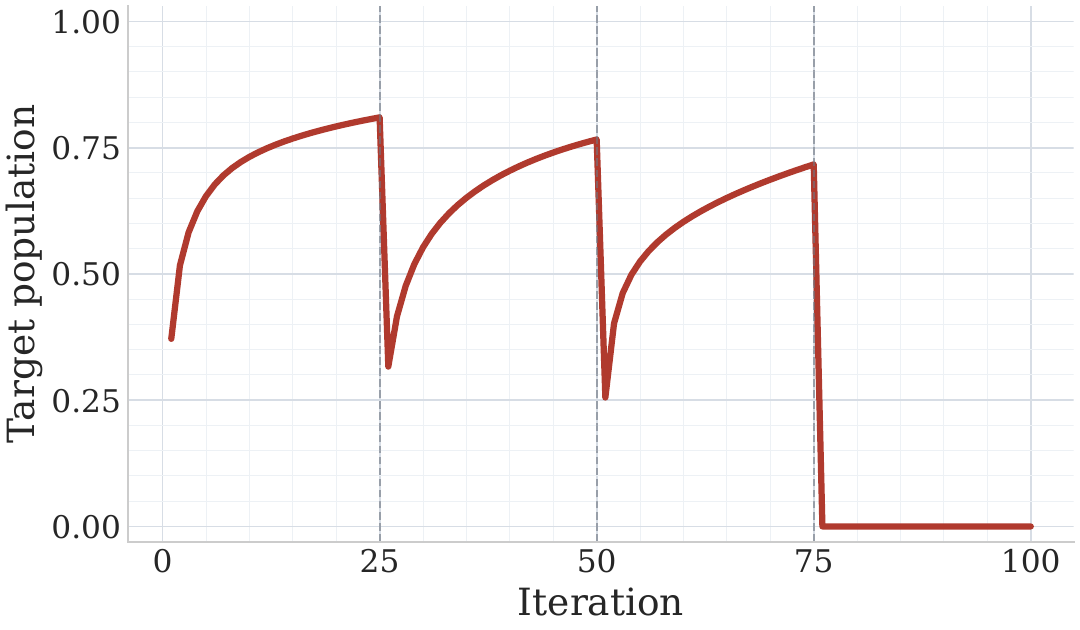}
        \caption{$\alpha = 5 \cdot 10^{-3}$, 25 iterations per stage}
        \label{fig:sweep_aggressive}
    \end{subfigure}

    \vspace{0.5cm}
    \begin{subfigure}[b]{0.48\textwidth}
        \includegraphics[width=\textwidth]{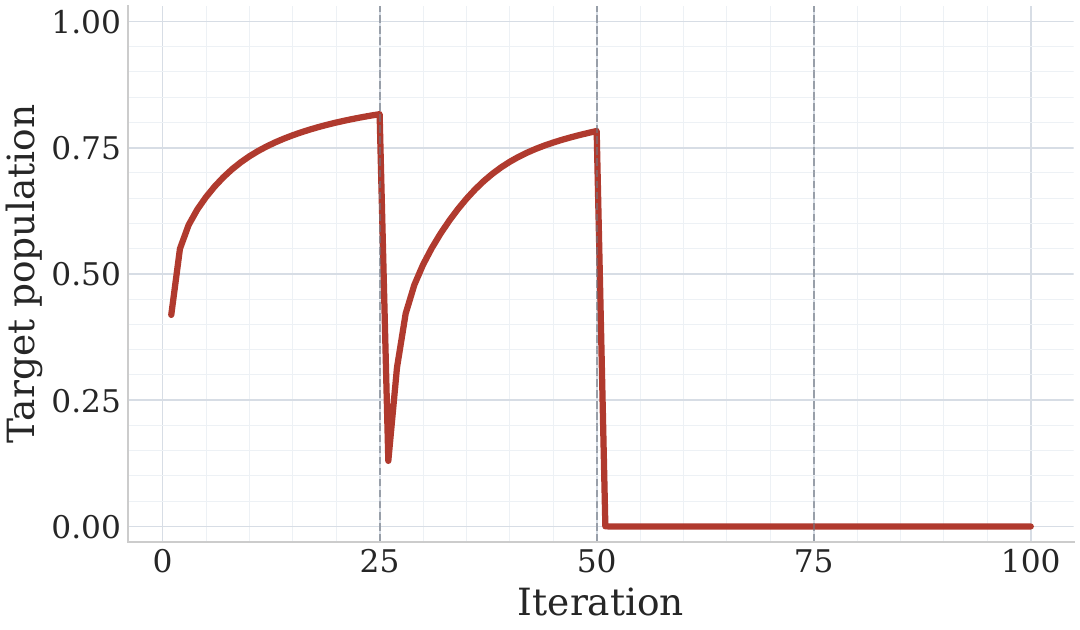}
        \caption{$\alpha = 2 \cdot 10^{-2}$, 25 iterations per stage}
        \label{fig:sweep_weak}
    \end{subfigure}
    \hfill
    \begin{subfigure}[b]{0.48\textwidth}
        \includegraphics[width=\textwidth]{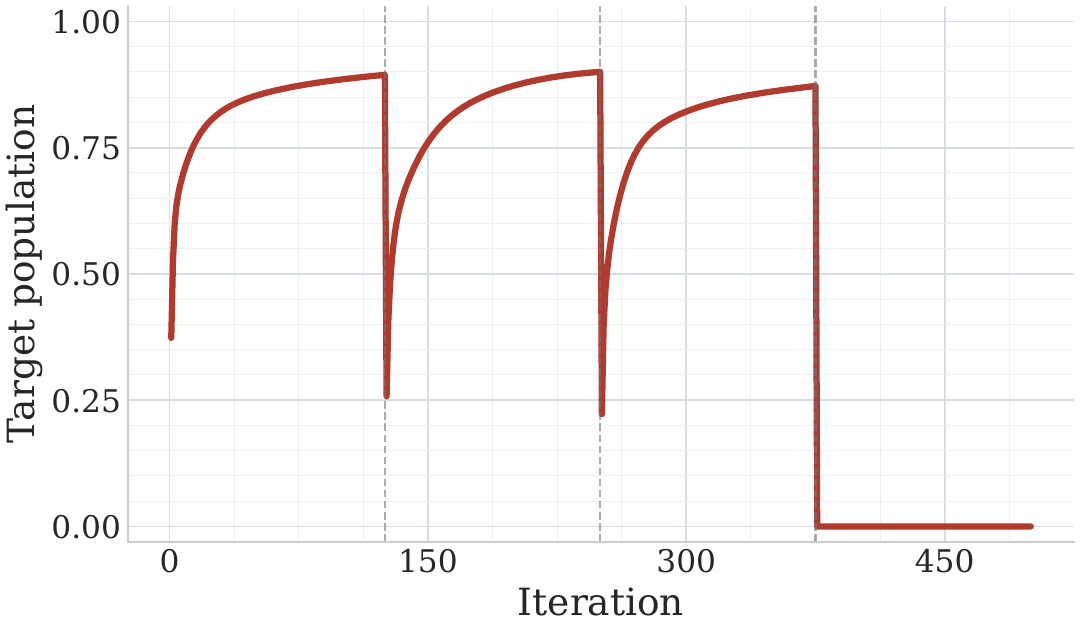}
        \caption{$\alpha = 10^{-2}$, 125 iterations per stage}
        \label{fig:sweep_overopt}
    \end{subfigure}

\caption{Evolution of the final target population $J_T$ during the four-stage sequential protocol for different penalty parameters and numbers of iterations per stage. Panels (a)-(d) correspond to the four tested configurations; vertical changes in the curves mark changes of intermediate target between stages.}
\label{fig:parameter_sweep}
\end{figure}
\begin{figure}[htbp]
        \centering
        \includegraphics[width=.65\textwidth]{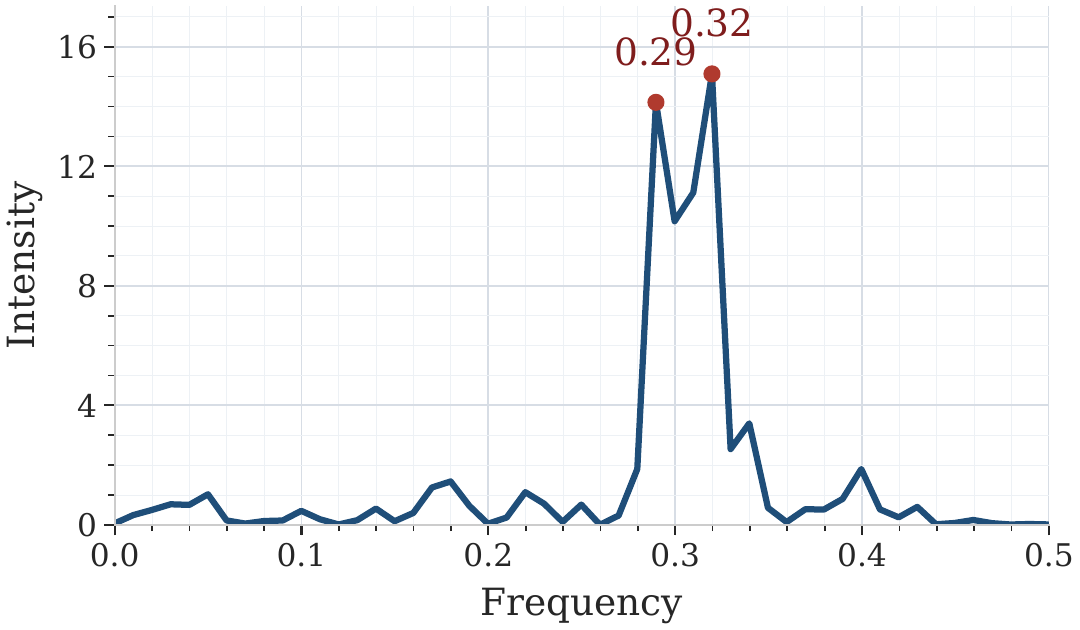}
    \caption{Power spectrum of the optimized control fields at the end of last stage of the sequential protocol. The frequency resolution is $\Delta f \approx 0.01$.}   
    \label{fig:fourier_spectra}
\end{figure}
\begin{figure}[htbp]
    \centering
    \begin{subfigure}[b]{0.48\textwidth}
        \centering
        \includegraphics[width=\textwidth]{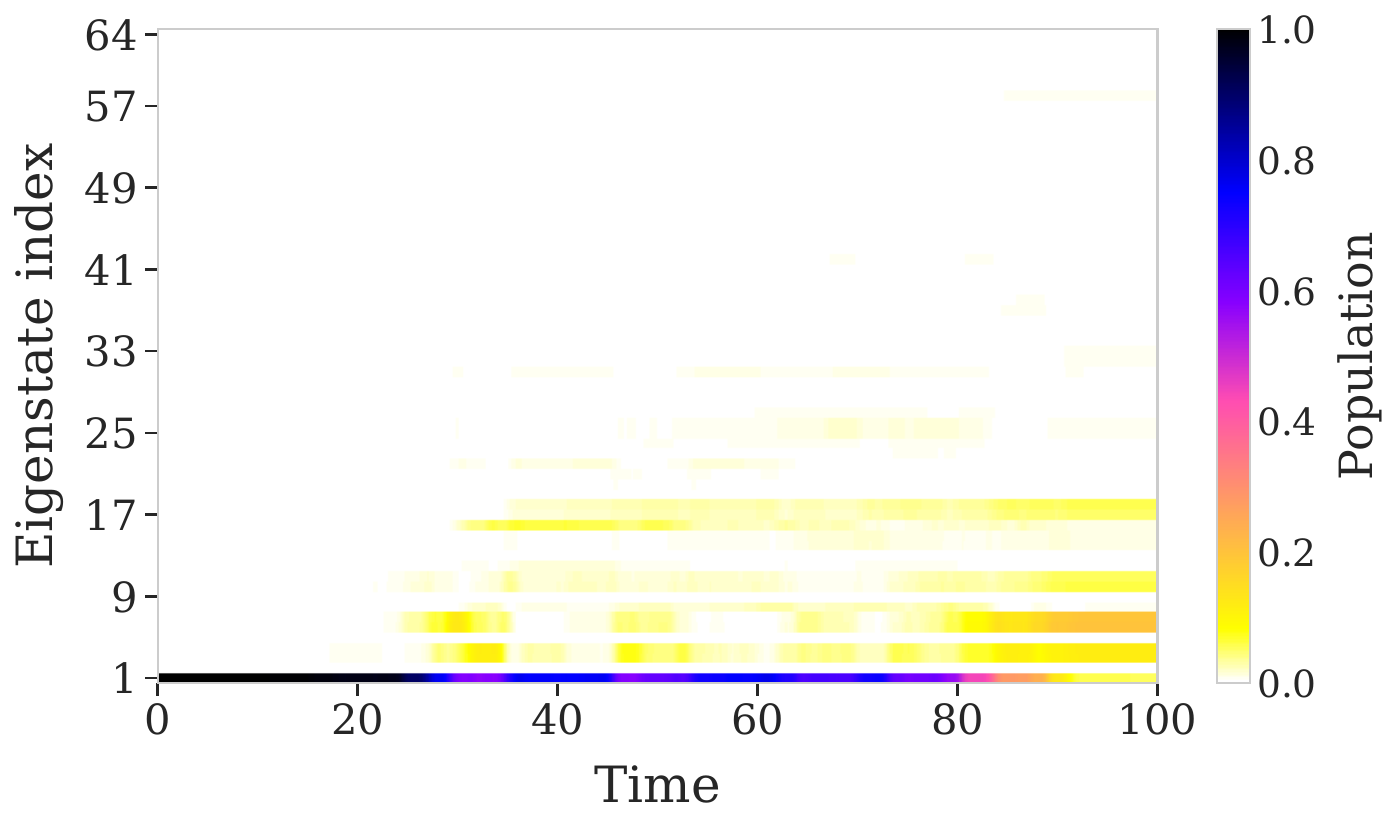}
        \caption{Stage 1, iteration 25}
        \label{fig:eig_25}
    \end{subfigure}
    \hfill
    \begin{subfigure}[b]{0.48\textwidth}
        \centering
        \includegraphics[width=\textwidth]{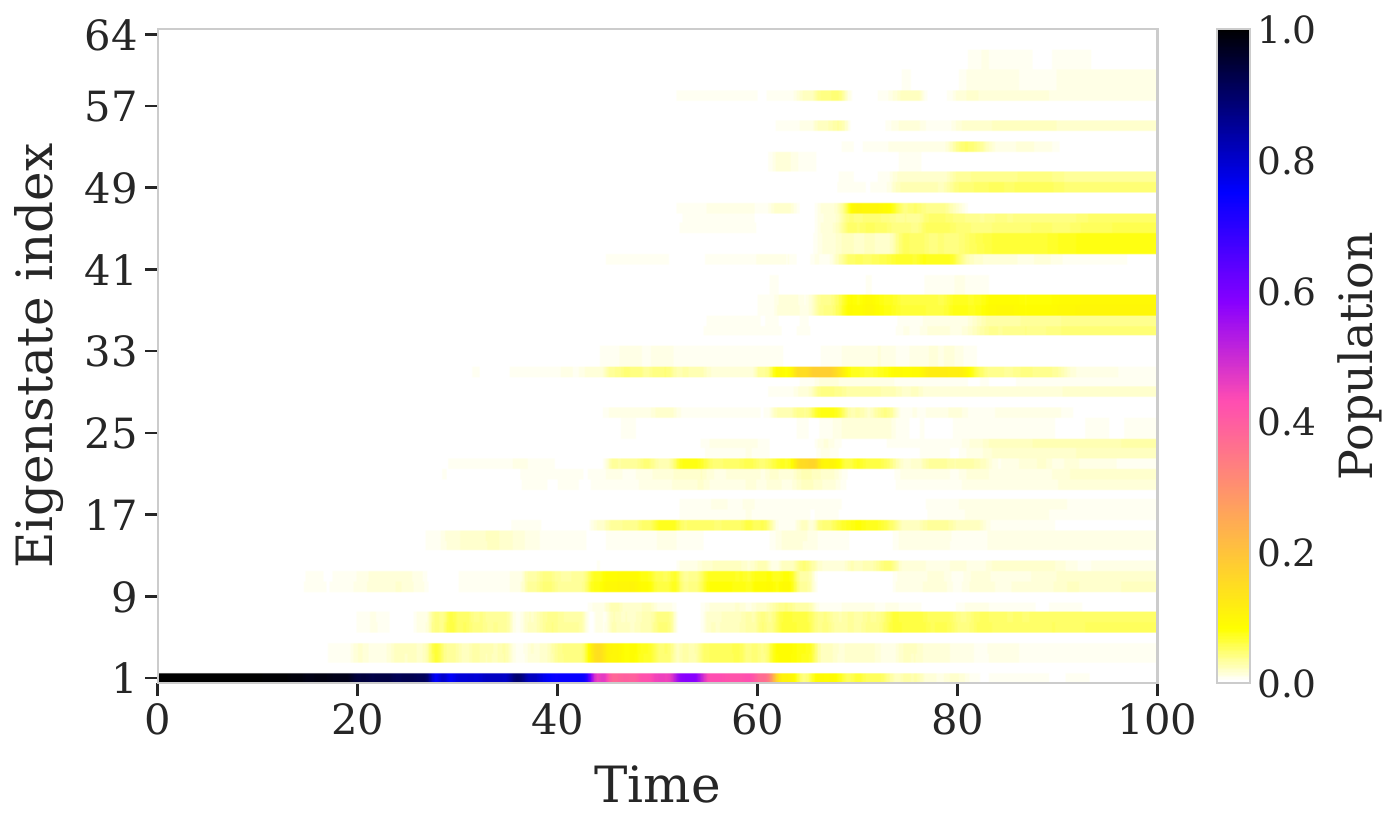}
        \caption{Stage 2, iteration 50}
        \label{fig:eig_50}
    \end{subfigure}

    \vspace{0.4cm}

    \begin{subfigure}[b]{0.48\textwidth}
        \centering
        \includegraphics[width=\textwidth]{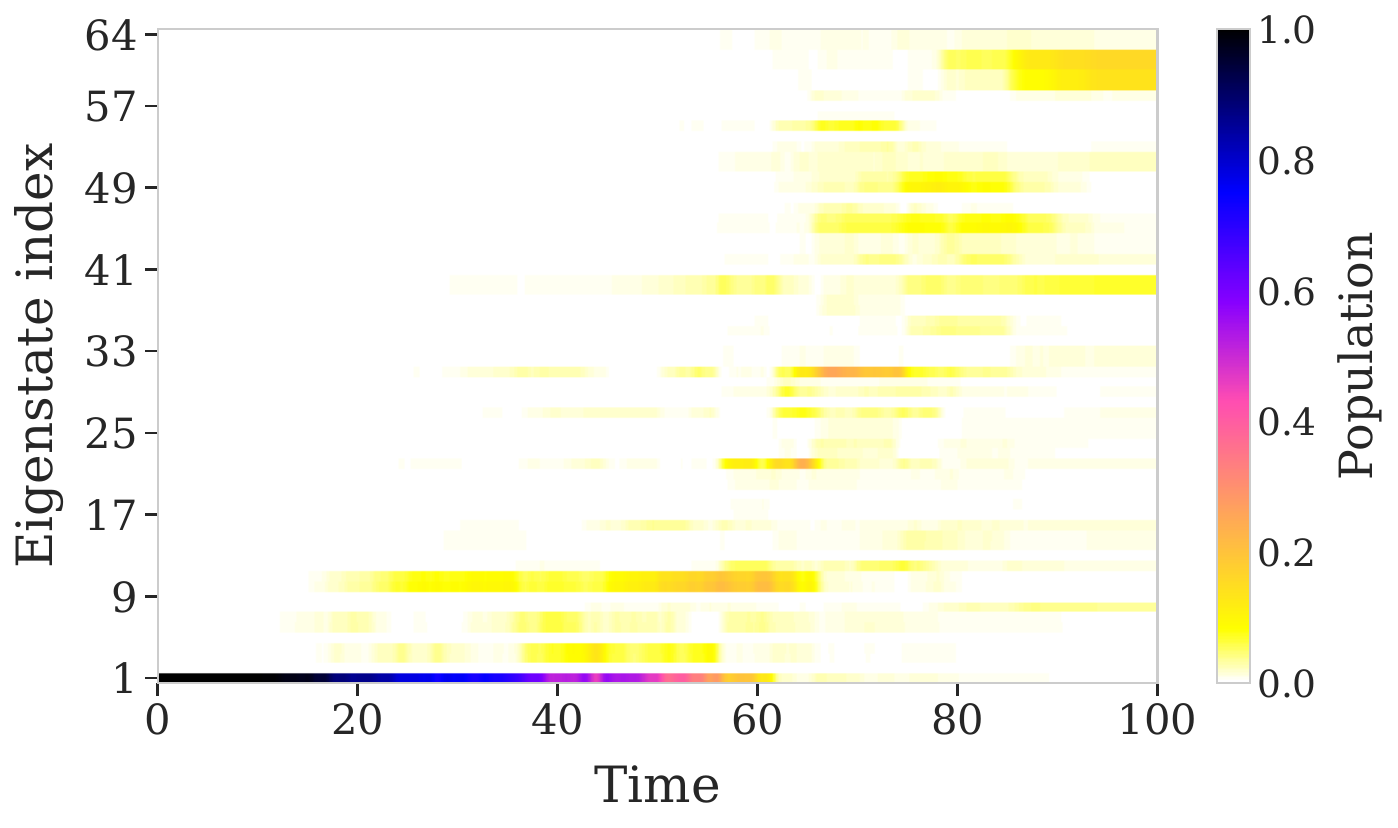}
        \caption{Stage 3, iteration 75}
        \label{fig:eig_75}
    \end{subfigure}
    \hfill
    \begin{subfigure}[b]{0.48\textwidth}
        \centering
        \includegraphics[width=\textwidth]{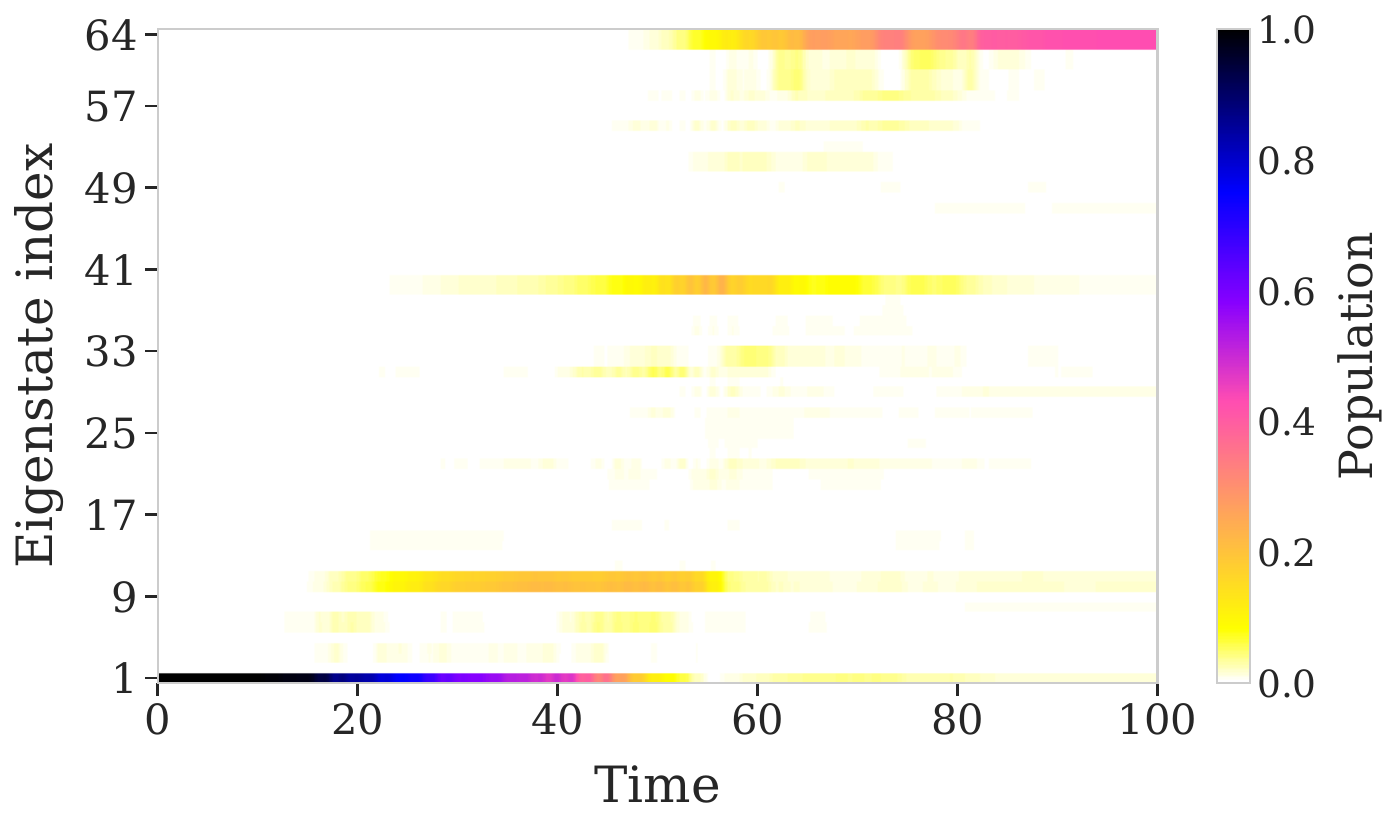}
        \caption{Stage 4, iteration 100}
        \label{fig:eig_100}
    \end{subfigure}
    \caption{Time evolution of the field-free eigenstate populations under the optimized field obtained at the end of each stage. Panels (a)-(d) correspond to stages 1-4, respectively; the color scale represents the instantaneous population of each eigenstate.
    }
    \label{fig:eigenstate_paths}
\end{figure}

To determine working conditions for the sequential protocol, four parameter configurations were tested. The stability of the algorithm depends on both the penalty parameter $\alpha$, and the number of iterations per stage. The four tested scenarios were:
\begin{itemize}
    \item 25 iterations per stage with $\alpha = 10^{-2}$.
    \item 25 iterations per stage with $\alpha = 5 \cdot 10^{-3}$.
    \item 25 iterations per stage with $\alpha = 2 \cdot 10^{-2}$.
    \item 125 iterations per stage with $\alpha = 10^{-2}$.
\end{itemize}
Convergence of this protocol to a non-trivial optimal solution is not guaranteed. Figure~\ref{fig:parameter_sweep} illustrates stage efficiency by tracking the target population at each iteration. Notably, maximizing the final observable at intermediate stages can cause the algorithm to collapse to a trivial zero field in the final stage, as in the case $\alpha = 10^{-2}$ with 125 iterations. Here, high intermediate target populations yield insufficient component populations for moderating overlaps at the end of the propagation time. On the other hand, excessively weak fields, $\alpha = 2 \cdot 10^{-2}$, fail to populate intermediate states required for targeting the previous stages. The main obstacle of this multistage protocol is therefore the precise set of parameters of the QOC algorithm as well as the choice of the intermediate target states.
\\
By setting $\alpha = 10^{-2}$, convergence to a non-trivial solution is guaranteed. Each stage successfully transfers population into adjacent sectors, providing adequate costate amplitude for the following stage to reach a final target population of $J_T = 0.854$. 
\\
\begin{figure}[htbp]
    \centering
    \begin{subfigure}[b]{0.49\textwidth}
        \centering
        \includegraphics[width=\textwidth]{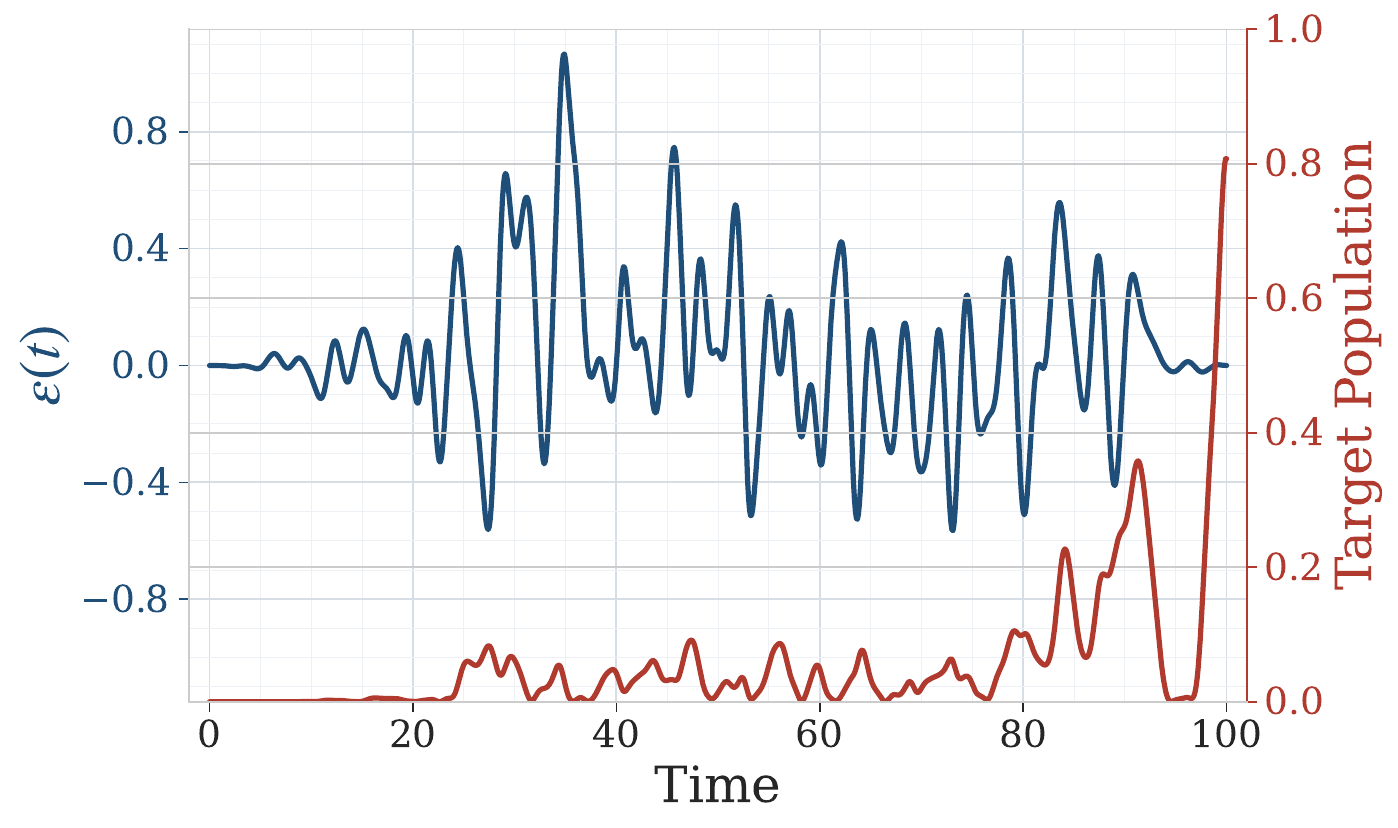}
        \caption{Stage 1, iteration 25}
        \label{fig:field_25}
    \end{subfigure}
    \hfill
    \begin{subfigure}[b]{0.49\textwidth}
        \centering
        \includegraphics[width=\textwidth]{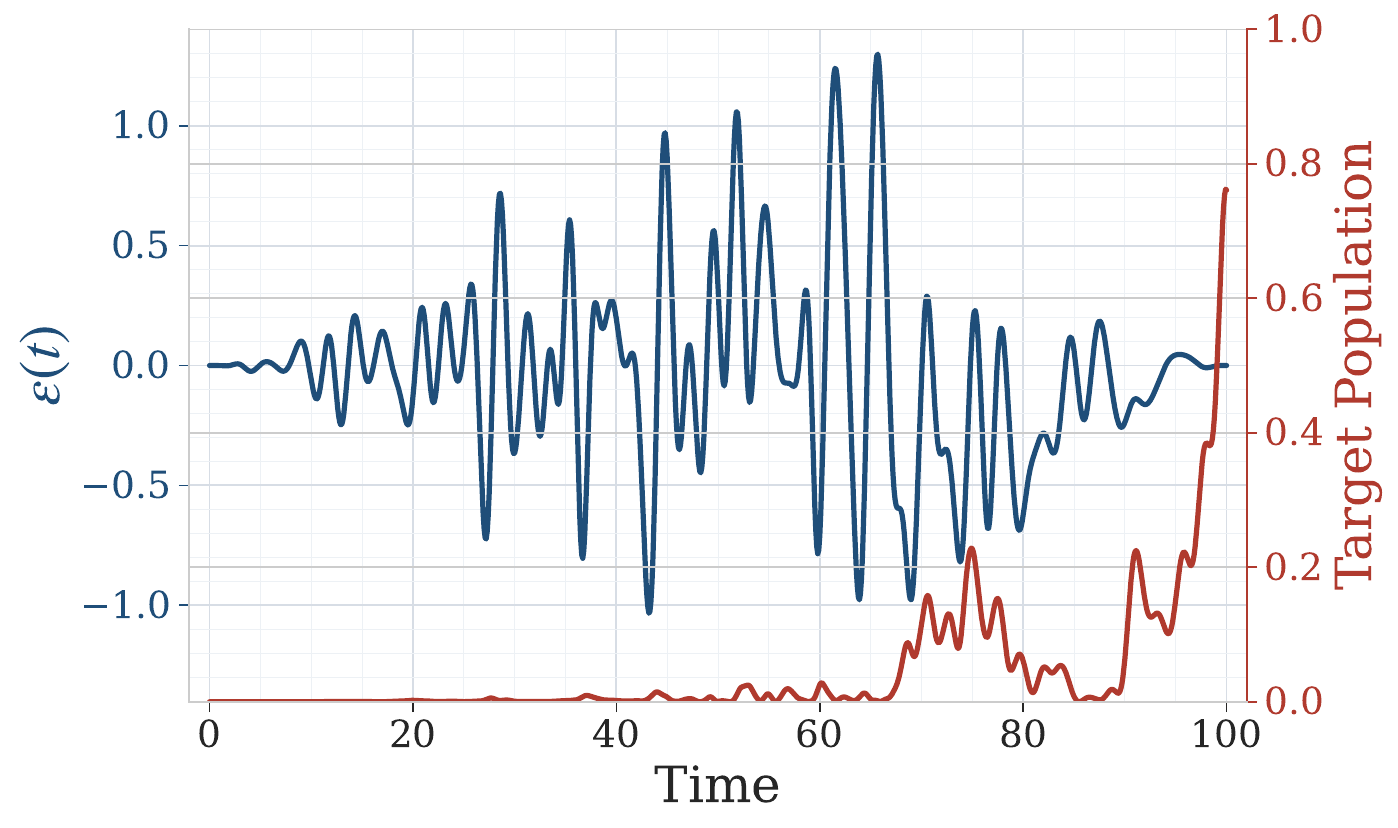}
        \caption{Stage 2, iteration 50}
        \label{fig:field_50}
    \end{subfigure}

    \vspace{0.4cm}

    \begin{subfigure}[b]{0.49\textwidth}
        \centering
        \includegraphics[width=\textwidth]{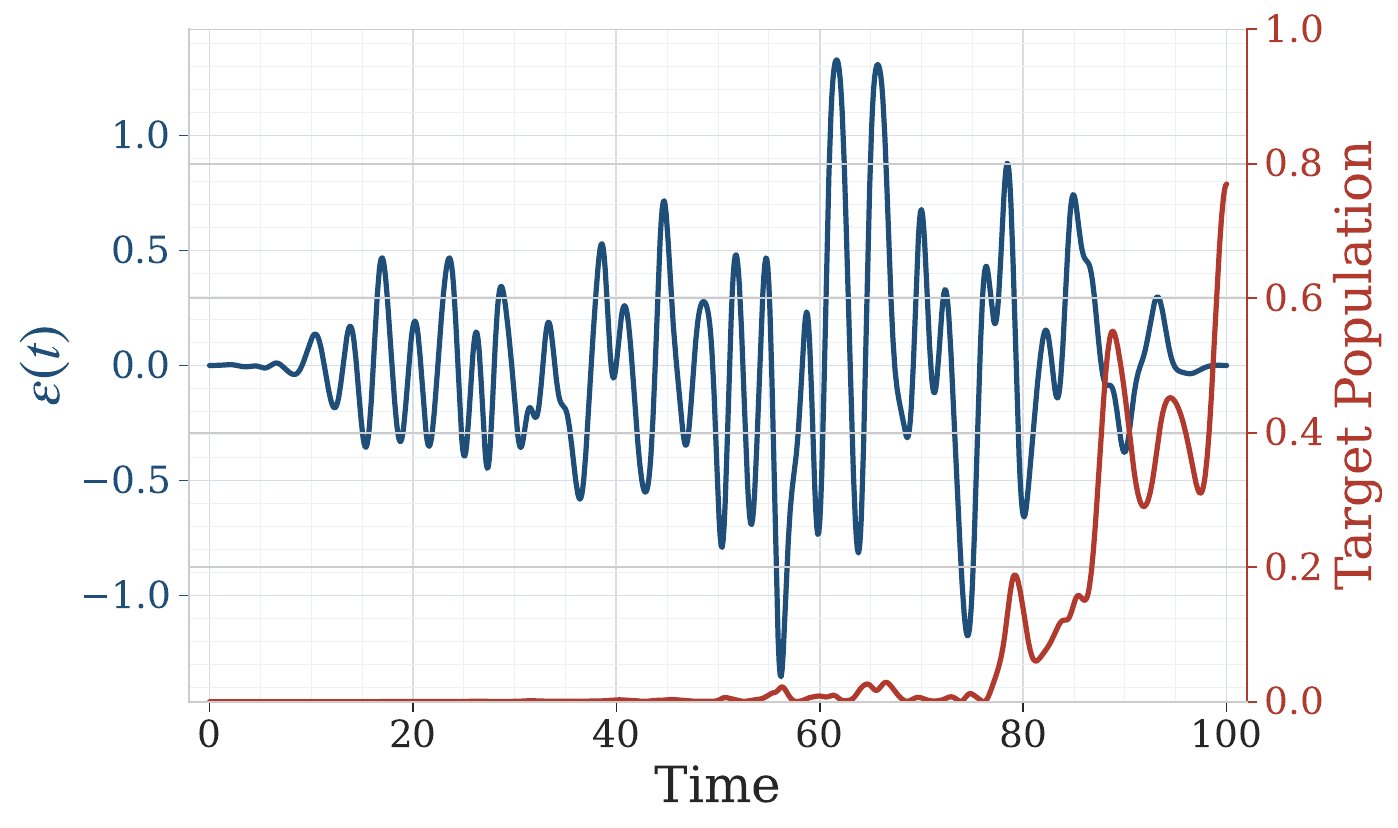}
        \caption{Stage 3, iteration 75}
        \label{fig:field_75}
    \end{subfigure}
    \hfill
    \begin{subfigure}[b]{0.49\textwidth}
        \centering
        \includegraphics[width=\textwidth]{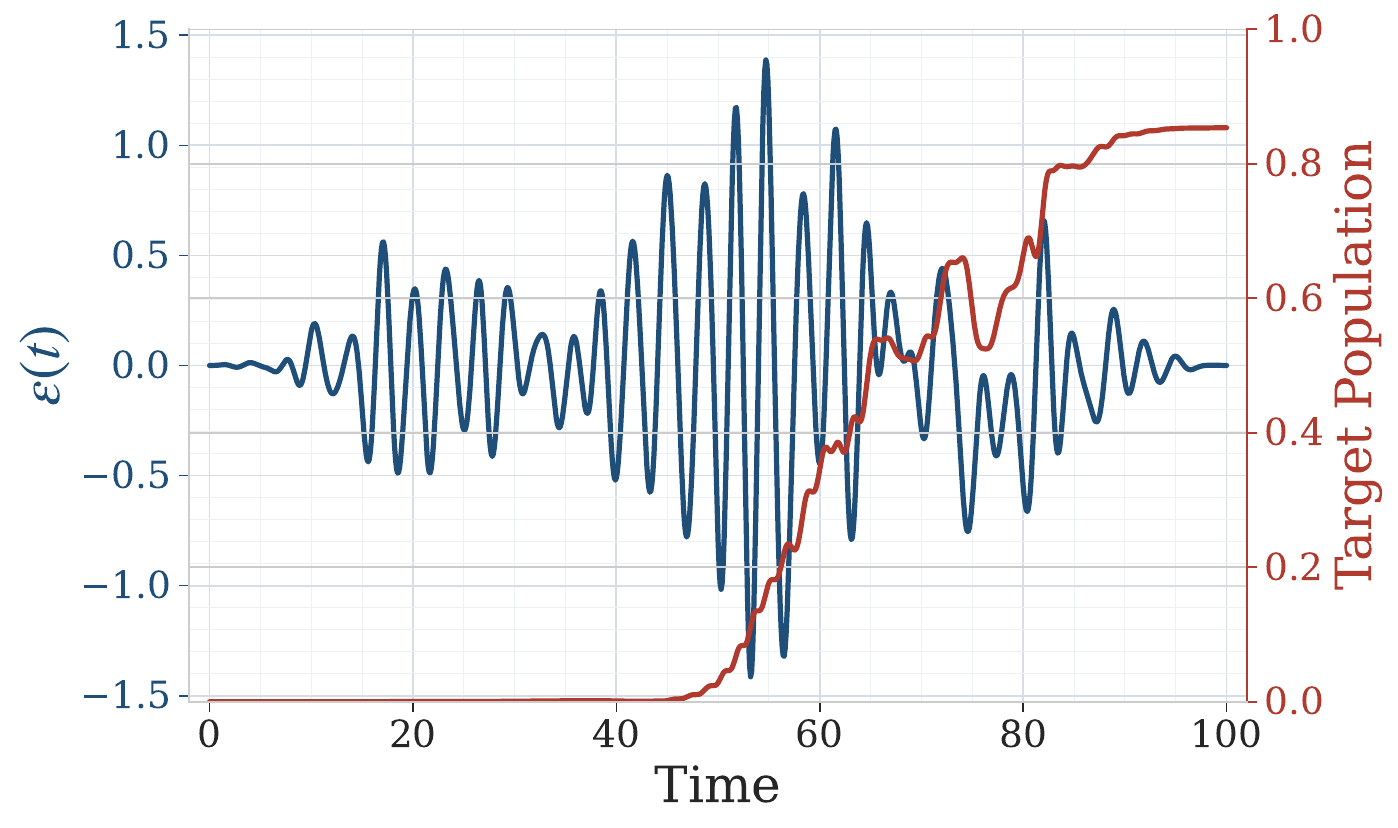}
        \caption{Stage 4, iteration 100}
        \label{fig:field_100}
    \end{subfigure}
    \caption{Optimized control field $\varepsilon(t)$ (blue) and time-dependent target population (red) for the final iteration of each stage of the sequential protocol. Panels (a)-(d) correspond to stages 1-4, respectively.}
    \label{fig:campos_observables}
\end{figure}
To understand the physical mechanism behind the multistage protocol, we analyze population transfer among eigenstates within the perturbative regime defined by weak field intensities. The fields for each stage and the final stage power spectrum are presented in Fig.~\ref{fig:campos_observables} and Fig.~\ref{fig:fourier_spectra}, respectively. Although the intermediate-stage fields are complex and lack obvious periodicity, three distinct temporal features can be distinguished in the last stage: an initial weak signal up to $t \approx 40$, an intermediate signal spanning $t \approx 40$ to $70$, and a final irregular perturbation. This dynamic arises because the intermediate targets (stages 1--3) are not eigenstates, in contrast to the GHZ target in stage 4, which is an eigenstate of the field-free Hamiltonian. We find that the power spectrum for the optimal field is characterized by an intense band from approximately $0.29$ to $0.32$, where we expect the relevant population transfer to occur.
\\
Now we turn to the population of each eigenstate, shown in Fig.~\ref{fig:eigenstate_paths}, and check the compatibility of the populations and the analyzed spectrum. The system initially resides entirely in the ground state [Eq.~\eqref{eq:ground_state_full}], characterized by zero net magnetization (three spins in state 0 and three in state 1). Around $t \approx 40$, the population flows into the degenerate eigenstates $|E_{10}\rangle$ and $|E_{11}\rangle$. These states represent the first step in the magnetization ladder. They are constructed as linear combinations of all basis states containing either two spins up ($|1\rangle$s) and four spins down ($|0\rangle$s), or four spins up ($|1\rangle$s) and two spins down ($|0\rangle$s), giving and absolute magnetization of $|S_{\mathrm{tot}}^z| = 1$. Due to the large number of terms, the following abbreviated expressions are used:
\begin{equation}
\begin{split}
    |E_{10}\rangle =\; & \phantom{+} 0.0277 |000011\rangle - 0.1611 |000101\rangle + \dots \\
                       & + 0.0036 |111010\rangle - 0.0006 |111100\rangle
\end{split}
\label{eq:eigenstate_10}
\end{equation}

\begin{equation}
\begin{split}
    |E_{11}\rangle =\; & - 0.0006 |000011\rangle + 0.0036 |000101\rangle - \dots \\
                       & + 0.1611 |111010\rangle - 0.0277 |111100\rangle
\end{split}
\label{eq:eigenstate_11}
\end{equation}

Subsequently, near $t \approx 60$, the population transfers to the degenerate eigenstates $|E_{39}\rangle$ and $|E_{40}\rangle$. These states are a superposition of basis states with either one $|1\rangle$ and five $|0\rangle$s, or five $|1\rangle$s and one $|0\rangle$, showing a further jump in polarization to $|S_{\mathrm{tot}}^z| = 2$.\\
\begin{equation}
\begin{split}
    |E_{39}\rangle =\; & - 0.2858 |011111\rangle + 0.5033 |101111\rangle \\
    &+ 0.4061 |110111\rangle - 0.4061 |111011\rangle\\
                       & - 0.5033 |111101\rangle + 0.2858 |111110\rangle
\end{split}
\label{eq:eigenstate_39}
\end{equation}

\begin{equation}
\begin{split}
    |E_{40}\rangle =\; & - 0.2858 |000001\rangle + 0.5033 |000010\rangle \\
    &+ 0.4061 |000100\rangle - 0.4061 |001000\rangle\\
    &- 0.5033 |010000\rangle + 0.2858 |100000\rangle
\end{split}
\label{eq:eigenstate_40}
\end{equation}
Finally, in the last segment of the propagation, the system reaches the target state, composed of the highest energy eigenstates $|E_{63}\rangle$ and $|E_{64}\rangle$. These correspond to the states of total polarization of $|S_{\mathrm{tot}}^z| = 3$.\\
\begin{equation}
\begin{split}
    |E_{63}\rangle =\; & |000000\rangle \\
    |E_{64}\rangle =\; & |111111\rangle
\end{split}
\label{eq:eigenstate_63_64}
\end{equation}
The sequence of transitions ($|S_{\mathrm{tot}}^z| = 0 \to 1 \to 2 \to 3$) reflects a discrete and progressive increase in absolute total magnetization. Since the control field is local and acts only on the first spin, the excitation propagates through the chain and increases the absolute magnetization step-by-step.\\
This stepwise change in absolute magnetization is consistent with the transition energies between these eigenstates and the power spectrum in Fig.~\ref{fig:fourier_spectra}. The eigenvalues involved are $E_1 \approx -3.3944$, $E_{10/11} \approx -1.4009$, $E_{39/40} \approx 0.6196$, and $E_{63/64} = 2.5$. The required angular frequencies ($f = \Delta E / 2\pi$) for the sequential transitions are listed below:
\begin{itemize}
    \item $1 \to 10/11$, $\Delta E = 1.9935 \implies f \approx 0.317$
    \item $10/11 \to 39/40$, $\Delta E = 2.0205 \implies f \approx 0.322$
    \item $39/40 \to 63/64$, $\Delta E = 1.8804 \implies f \approx 0.299$
\end{itemize}
All three transition frequencies fall within the $0.29 - 0.32$ range, in agreement with the frequency band observed in the Fourier spectrum. Let us remark that a direct transition from the initial ground state to the highly magnetized target state would require an energy jump of $\Delta E_{\text{direct}} = 2.5 - (-3.3944) = 5.8944$, which corresponds to a frequency of $0.938$. This high frequency component is absent from the spectrum, supporting the interpretation that the local control does not generate a direct global multi-spin flip.\\
Spectral analysis of control fields driving transition dynamics, reveals dominant resonant frequencies matching the multi-step pathway $\ket{E_1} \to \ket{E_{10/11}} \to \ket{E_{39/40}} \to \ket{E_{63/64}}$. Utilizing these extracted transition frequencies, seed pulses constructed with time-localized frequency components accelerate convergence.

\subsection{Comparison with Time-Localized Signals Seed}
\label{sec:time_localized_signals_seed}
\begin{figure}[htbp]
    \centering
    \begin{subfigure}[b]{0.49\textwidth}
        \centering
        \includegraphics[width=\textwidth]{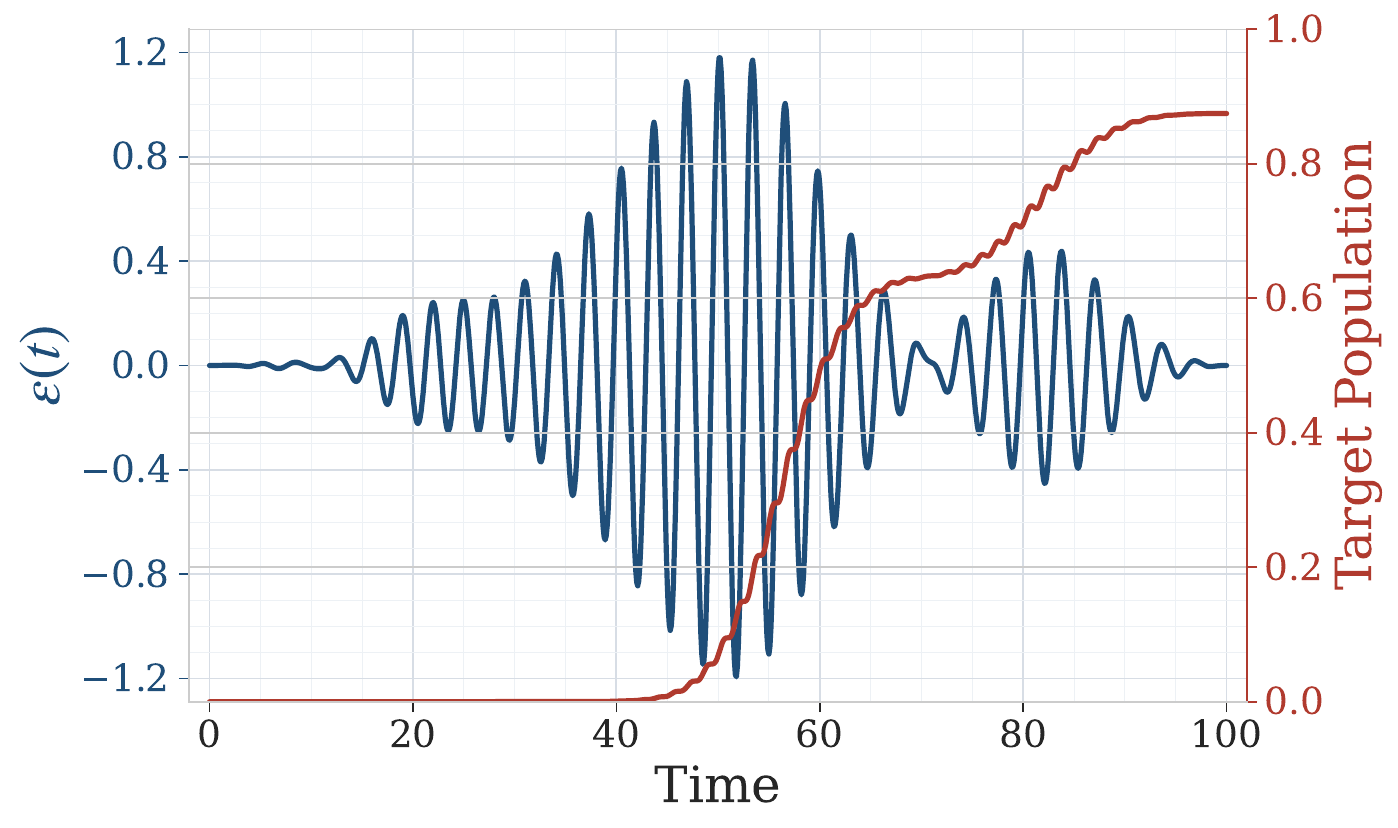}
        \caption{Optimal field for simultaneous seed.}
    \end{subfigure}
    \hfill
    \begin{subfigure}[b]{0.49\textwidth}
        \centering
        \includegraphics[width=\textwidth]{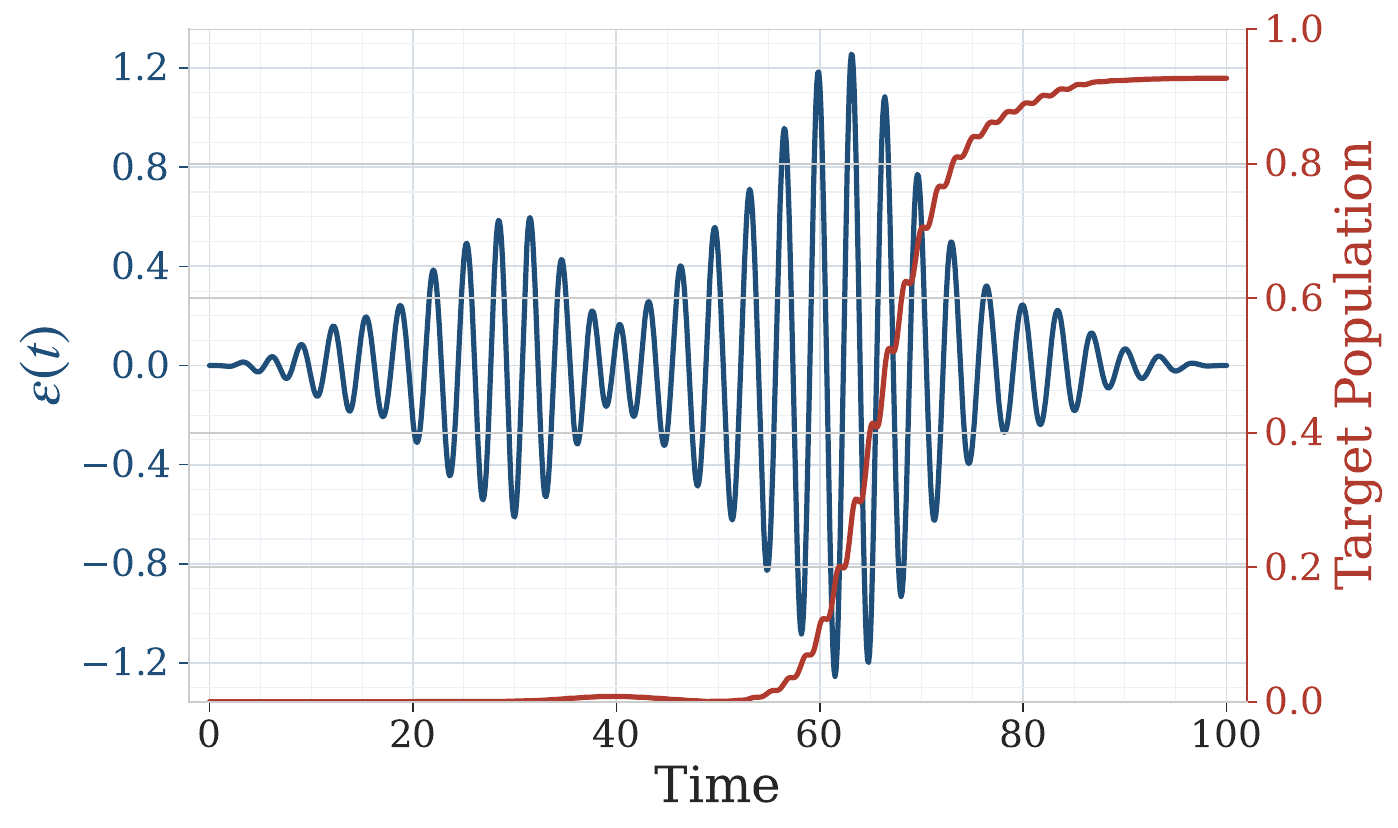}
        \caption{Optimal field for sequential seed.}
    \end{subfigure}
    \caption{\label{fig:time_localized_seed} Optimal fields obtained for predesigned seeds (blue) and the target population (red) as a function of time.}
\end{figure}

The results of Sec.~\ref{sec:sequential_protocol} show that arbitrary superpositions require a complex seed field to generate sufficient initial population. For an eigenstate target such as the fully polarized GHZ state, the optimized dynamics follows a simpler sequence of transitions. As established in Sec.~\ref{sec:sequential_protocol}, the path from the zero magnetization ground state to the maximum magnetization target given by the multistage protocol, proceeds through intermediate eigenstates ($|E_1\rangle \to |E_{10/11}\rangle \to |E_{39/40}\rangle \to |E_{63/64}\rangle$).\\
Because the stepwise transition mechanism is governed by transitions among eigenstates, the corresponding optimized control field mainly contains a set of well-defined frequencies. By designing a seed field for the GHZ state constructing an initial seed field that combines the frequencies corresponding to all the transitions involved, the optimization algorithm can be guided from the first iteration. The objective of this tailored approach is to increase the final target population.
\\
To test this approach, the seed is constructed from the three transition frequencies identified previously ($f_1 \approx 0.317$, $f_2 \approx 0.322$, and $f_3 \approx 0.299$) but varying their temporal envelopes. All components use an amplitude of $A_i = 0.35$.
\\
The first design sums the three frequencies weighted by a sine-squared envelope over the standard propagation window $T = 100$, acting in simultaneously:
\begin{equation}
    \varepsilon^{(0)}(t) = \sin^2\left(\frac{\pi t}{T}\right) \sum_{i=1}^{3} A_i \cos(2\pi f_i t).
    \label{eq:seed_simultaneous}
\end{equation}
\\
The second design mimics the sequential change $\Delta |S_{\mathrm{tot}}^z| = 1$ by characterizing the signals with gaussian envelopes located at different times $t_i$ with different widths $\sigma_i$:
\begin{equation}
    \varepsilon^{(0)}(t) = \sum_{i=1}^{3} A_i \cos(2\pi f_i t) \exp\left[-\frac{1}{2}\left(\frac{t-t_i}{\sigma_i}\right)^2\right] \ ,
    \label{eq:seed_chronological}
\end{equation}
where we set Gaussian envelopes centered at $t_1 = 25$, $t_2 = 50$, and $t_3 = 75$ with the standard deviations $\sigma_1=10$, $\sigma_2=12$ and $\sigma_3=18$, respectively. The optimal fields obtained for each designed seed field are shown in Fig.~\ref{fig:time_localized_seed} along with the population of the GHZ state as a function of time for the 20th iteration. 
\\
First, we realize that the observe dynamics are different for each designed seed field. For the simultaneously localized seed, we observe that the overlap between the first two transitions is large so that the population transfer $\ket{E_1}\rightarrow \ket{E_{10/11}}$ and $\ket{E_{10/11}}\rightarrow \ket{E_{39/40}}$ occurs consecutively. Hence, the population rises smoothly up to 0.6 at $t\approx 65$. The population of the GHZ state remains constant up to approximately $t\approx 75$, caused by the isolated signal corresponding to the last transition $\ket{E_{39/40}}\rightarrow \ket{E_{63/64}}$. However, the sequential seed field converges to a continuous transition process $\ket{E_1}\rightarrow\ket{E_{10/11}}\rightarrow \ket{E_{39/40}}\rightarrow \ket{E_{63/64}}$.
\\
Finally, we find that the obtained $J_T$ is larger than in the multistage protocol, specifically $0.875$ and $0.927$ for the simultaneously and sequential frequency seeds. We summarize in Table~\ref{tab:ghz_summary} the optimal values obtained. Let us remark that we may achieve a higher fidelity by increasing the time of the process. For instance, for $T=200$ the fidelity is higher than $0.982$ for 20 iterations.

\begin{table}[h]
\caption{Performance summary for GHZ target state preparation.}
\label{tab:ghz_summary}
\begin{tabular}{lccc}
\toprule
Seed Protocol & $T$ & Iterations & Final $J_T$ \\
\midrule
Eq.~\ref{eq:hahn_seed} & 100 & 20 & $< 10^{-4}$ \\
Sequential 4-Stage & 100 & 100 & 0.854 \\
Simultaneous Frequency Seed & 100 & 20 & 0.875 \\
Sequential Frequency Seed & 100 & 20 & 0.927 \\
\bottomrule
\end{tabular}
\end{table}

\section{Conclusion}
\label{sec:conclusions}
We have investigated the quantum optimal control of an open-boundary anisotropic XXZ six-spin chain driven solely by a single local transverse magnetic field applied to a terminal spin. Our results demonstrate that a single local actuator can leverage internal exchange interactions to steer the global spin array and prepare highly entangled target states, such as the Greenberger--Horne--Zeilinger state, with fidelities exceeding $0.927$. We show that excitation transfer across large magnetization distances avoids direct multi-spin flips, propagating instead via a sequential transition cascade through intermediate energy eigenstates in adjacent magnetization sectors. Fourier spectra of the optimized fields confirm that this population transfer is mediated by resonant frequency components matching the internal energy gaps along the multi-step pathway.

From an algorithmic perspective, we find that gradient-based methods such as Krotov's algorithm suffer from numerical stagnation when targeting high-magnetization states from simple seed fields due to the small-overlap mechanism in costate updates used to update the control field. We resolve this optimization bottleneck by implementing a multi-stage sequential target protocol that gradually maximizes population in states of increasing $|S^z|$. By using the optimized pulse from each stage as the seed field for the next, we obtain large costate overlap to ensure convergence and prevent trivial zero-field solutions. This multi-stage trajectory reveals a sequence of three well-defined transitions, enabling us to design time-localized frequency seeds. Overall, this demonstrates that minimal local control coupled with spectrally tailored initialization offers a robust, high-fidelity scheme for quantum state engineering in constrained spin networks.

\begin{backmatter}
\bmhead{Acknowledgments}
This research was supported by Grants No. PID2019-105458RB-I00, PID2021-122796NB-I00, PID2021-122839NB-I00 and PID2022-138288NB-C33, funded by the MICIU/AEI/10.13039/501100011033 and the ERDF/EU. The MATRIX-CM project is also aknowledged.

\section*{Declarations}
\bmhead{Conflict of interest}
The authors declare that they have no conflict of interest.

\bmhead{Data Availability}
Data supporting the findings of this study are available from the corresponding author upon reasonable request.
\end{backmatter}

\bibliography{thesis}

\end{document}